# PXIE: PROJECT X INJECTOR EXPERIMENT[*]

S. Nagaitsev[#], V. Lebedev, S. Holmes, R. Kephart, J. Kerby, S. Mishra, A. Shemyakin,
N. Solyak, R. Stanek, V. Yakovlev, Fermilab, Batavia, IL 60510, USA
P. N. Ostroumov, ANL, Argonne, IL 60439, USA
D. Li, LBNL, Berkeley, CA 94720, USA

*Abstract*

A multi-MW proton facility, Project X, has been proposed and is currently under development at Fermilab. We are planning a program of research and development aimed at integrated systems testing of critical components comprising the front end of the Project X linac at Fermilab. This program is being undertaken as a key component of the larger Project X R&D program. The successful completion of this program will validate the concept for the Project X front end, thereby minimizing the primary technical risk element within Project X. Integrated systems testing, known as the Project X Injector Experiment (PXIE), will be completed over the period FY12-16.

PXIE will include an H- ion source, a CW 2.1-MeV RFQ and two SC cryomodules providing up to 30 MeV energy gain at an average beam current of 1 mA (upgradable to 2 mA). Successful systems testing will also demonstrate the viability of novel front end technologies that will find applications beyond Project X in the longer term.

## PROGRAM GOALS

Project X is proposed as a multi-functional proton accelerator complex aimed to support Fermilab's leading role in the intensity frontier research over next several decades [1]. It will include a 3 GeV, 1 mA CW linac and a pulsed linac accelerating protons to 8 GeV for their further injection into the existing Main Injector synchrotron, as well as various high-power beam targets. It should replace the existing 40-year old injector complex and support a large number of experiments in the energy range from 1 to 120 GeV. The 3-GeV CW linac is the centerpiece of the facility. It has to deliver a proton beam to several experiments quasi-simultaneously with a beam structure which can be adjusted to each experiment's needs. This will be achieved by RF deflection of the beam after acceleration to 3 GeV. The time structure required by each experiment will be obtained by removing undesired bunches from the 162 MHz RFQ bunch stream in the medium energy beam transport (MEBT) at beam energy of 2.1 MeV [2]. While the use of the rf deflection to support multiple experiments has already been demonstrated at the CEBAF facility at Jefferson Lab, the wideband chopper is a unique device currently beyond state of the art.

Project X technical risks are primarily associated with the low energy part of the complex. The purpose of PXIE is to demonstrate that the technologies selected for the Project X front end can indeed meet the performance requirements established in the Reference Design, thereby mitigating the primary technical risk element associated with Project X. This goal would ideally be achieved in advance of Project X construction so that results can be properly reflected in the final Project X machine.

Specific technical PXIE program goals are to demonstrate: (1) reliable operation of a CW 2.1 MeV RFQ accelerator, (2) a bunch-by-bunch chopper, (3) low-$\beta$ acceleration in SC cryomodules, (4) a sufficiently small emittance growth in the course of acceleration and (5) good particle extinction for the removed bunches. PXIE has to operate at full Project X design parameters and deliver up to 1 mA average current while accommodating up to 100% chopping of 5 mA RFQ beam. The beam current upgradability requirement (to 2 mA) is determined by possible staging of the Project X and its future upgrades.

The PXIE design and construction is being carried out by a collaboration between Fermilab, ANL, LBNL, SLAC and Indian institutions. It is planned to have PXIE operational by the end of 2016.

## PXIE SUBSYSTEMS

Figure 1 shows the schematic layout of PXIE. The ion source supplies a 5-mA, 30-keV H- beam. The ion source is the TRIUMF-type DC volume-cusp negative hydrogen ion source fabricated by D-Pace, Inc. [3]. It is capable of generating a 15-mA DC beam, while all LEBT, RFQ and MEBT components are designed to support 10 mA. The emittance measurements performed in the LBNL yielded that the beam normalized rms emittance is within 0.09-0.12 mm mrad in the current range 2 to 10 mA. The ion source lifetime is about 300 hours. Two ion sources are planned to be used in Project X to achieve high reliability of its injector. An ion source selection will be performed with a switching dipole. Only one ion source will be used in PXIE.

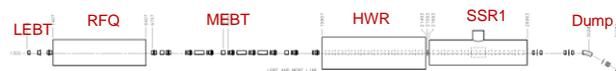

**Figure 1**: PXIE schematic layout. The total facility length is about 40 m.

The low energy beam transport (LEBT) section (~2 m long) delivers the beam to the RFQ. It also serves as a differential pumping section to prevent the gas flow from the ion source to the RFQ and includes the LEBT chopper which would form short beam current pulses (1-100 μs) with 60 Hz repetition rate for the machine commissioning. The LEBT optics relies partially on beam neutralization by residual gas ions and uses solenoids to focus the beam at the RFQ entrance.

---
* This work was supported by the U.S. DOE under Contract No. DE-AC02-07CH11359
# nsergei@fnal.gov

Figure 2 shows the RFQ conceptual design. It is a four-vane, 4.4 m long CW RFQ with a resonant frequency of 162.5 MHz, developed by LBNL [4]. Table 1 presents its main parameters. The RFQ final energy is chosen to be below the neutron production threshold in copper. At the same time it has to be large enough to reduce the beam space charge effects on the beam transport in the MEBT.

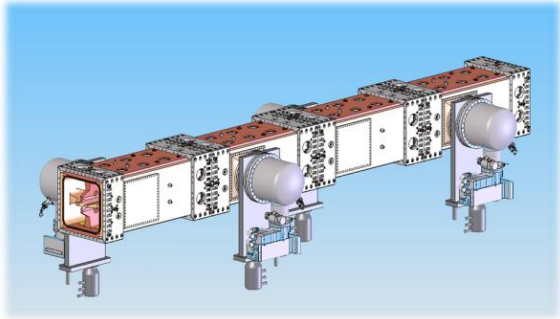

**Figure 2:** The RFQ conceptual design

**Table 1:** The RFQ parameters

| Parameter | Value | Units |
|---|---|---|
| Ion type | H- | |
| Beam current (nominal) | 5 | mA |
| Beam current (range) | 1 – 10 | mA |
| Trans. emitt. (rms, norm) | <0.25 | μm |
| Long. emitt. (rms) | 0.8-1.0 | keV-ns |
| Input energy | 30 | keV |
| Output energy | 2.1 | MeV |
| Duty factor | 100 | % |
| Frequency | 162.5 | MHz |

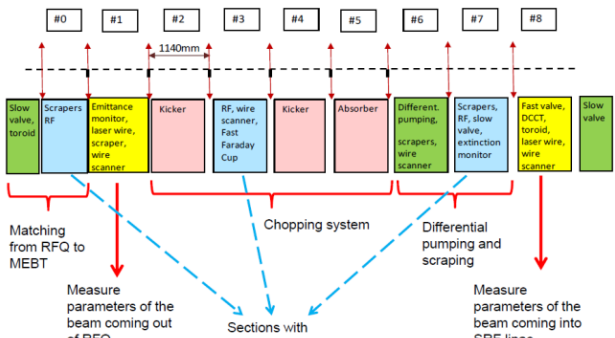

**Figure 3:** The conceptual MEBT schematic.

The PXIE MEBT serves the following functions:
- Forms the bunch structure required by the Project X users [2];
- Matches optical functions between the RFQ and the SRF cavities;
- Includes tools to measure the properties of the beam coming out of the RFQ and transported to the SRF cavities;
- Plays a role in a machine protection system.

Figure 3 shows the conceptual MEBT schematic. The MEBT has a periodic transverse focusing structure comprised of 9 drift sections (65 cm each) separated by quadrupole triplets. The overall length of the MEBT is about 10 m. Each drift section serves a distinct purpose as described in Fig. 3. Several main ideas drive such a concept. First, to reduce the kick voltage, the MEBT chopper employs two kickers separated by 180 deg. in betatron phase. Second, the RFQ frequency[*] was chosen sufficiently low so as to to reduce the kicker bandwidth to a manageable value (≤1 GHz). Third, the MEBT beam absorber [5] is being designed to absorb the full beam power, (21 kW), and, fourth, the differential pumping section isolates the absorber and the cryomodules to reduce the gas load to the cold section. The vacuum at the HWR cryomodule entrance was specified to be at or below $10^{-9}$ Torr to prevent potential performance degradation of the SC cavities because of high vacuum pressure. Figure 4 presents the transverse and longitudinal rms beam envelopes at a nominal beam current.

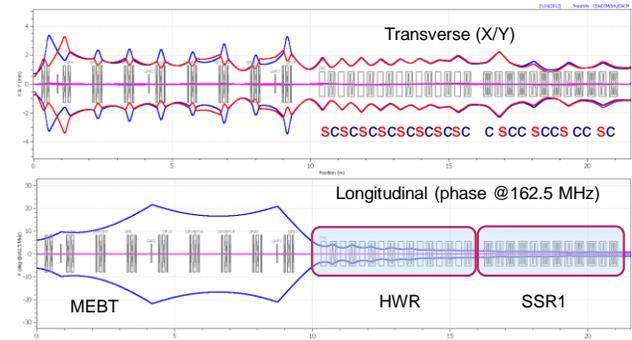

**Figure 4:** Transverse and longitudinal rms beam envelopes from the RFQ exit (2.1 MeV) to the exit of the SSR1 cryomodule (22 MeV). Peak beam current is 5 mA ($1.9 \times 10^8$ ppb).

The PXIE cryogenic section consists of two SC cryomodules, operating at 2K, separated by a warm section. The half-wave resonator (HWR) cryomodule [6] has 8 HW $\beta = 0.11$ cavities [7] operating at 162.5 MHz and separated by 8 superconducting focusing solenoids. Each solenoid accommodates a pair of transverse corrector coils and a BPM. The HWR CM (being designed by ANL) is ~5.9 m long and accelerates the beam from 2.1 MeV to ~11 MeV. The operational accelerating gradient was chosen to be 1.75 MV/cavity (at $\beta=0.11$). Figure 5 shows the HWR CM cavity and solenoid string assembly.

The single-spoke resonator (SSR1) cryomodule is being designed by FNAL [8]. It consists of 8 325-MHz single-spoke resonators ($\beta = 0.22$) [9] and 4 focusing solenoids, operating at 2K. Figure 6 shows the cavity and solenoid string assembly. Similar to the HWR CM, all solenoids will have corrector coils and BPMs. Its overall length (flange-to-flange) is about 5.3 m and it accelerates the beam from 11 to ~25 MeV. The operational accelerating gradient was chosen to be 2 MV/cavity (at β=0.22).

---
[*] Frequencies for PXIE rf cavities are chosen to be sub-harmonics of the ILC accelerating frequency of 1.3 GHz.

The high-energy test beam line (downstream of SSR1 CM) is designed to accommodate the beam diagnostics to measure the beam properties and the beam extinction for rf buckets emptied by MEBT chopper [10]. Finally, the PXIE beam dump at the end of the beam line is being designed for 50 kW. The dipole magnet immediately upstream of the beam dump will serve as a spectrometer to measure the beam energy. A variety of diagnostic tools will be installed throughout the PXIE beam line to support the commissioning and the R&D program [11].

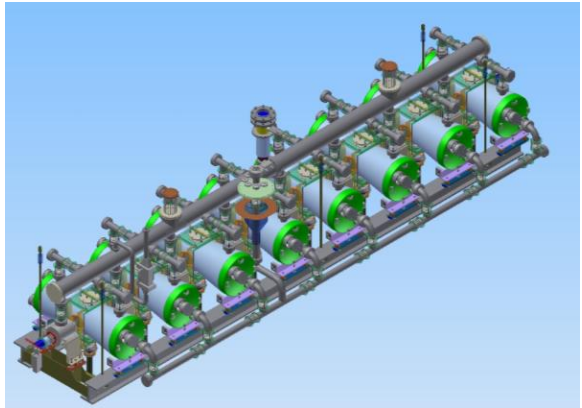

**Figure 5:** The HWR cavity and solenoid string assembly.

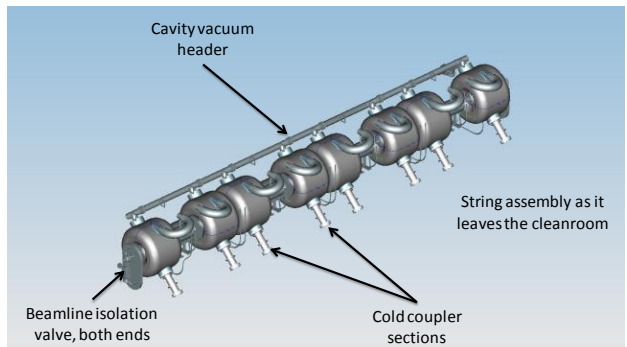

**Figure 6:** The SSR1 cavity and solenoid string assembly.

PXIE will be located in the existing Cryomodule Test Facility (CMTF) building which also will include an enclosure for ILC cryomodules testing. A cryo-plant supporting operation of both installations will have 500 W cooling power at 2 K and 4.1 kW at 40 K. It is expected that at normal operations PXIE will require cryo-power not exceeding 100 W at 2 K.

Many PXIE subsystems and components are being designed and constructed in collaboration with LBNL (LEBT and MEBT), SLAC (MEBT chopper driver) and Indian laboratories (high-level rf, beam instrumentation).

## SUMMARY

The PXIE program is being designed and constructed at Fermilab as the centerpiece of the Project X R&D program. It will provide an integrated systems test for Project X front end components and validate the concept for the Project X front end, thereby minimizing the primary technical risk element within the Reference Design.

Main technical issues to be addressed by PXIE are:
- LEBT: the beam neutralization, the chopper performance, and the beam stability;
- RFQ: the longitudinal halo formation and the high average power;
- MEBT: the beam dynamics, the chopper kicker and its driver, the absorber, the diagnostics, the extinction, and vacuum near SCRF cavities;
- HWR and SSR1: the losses, the beam acceleration, the effect of solenoids magnetic field on SC RF cavities, the microphonics;
- Beam line: the beam properties, the beam extinction, the beam losses and halo.